\documentclass[10pt,letterpaper,twocolumn]{article}
\usepackage[10pt,nocopyright]{sigmin}

\usepackage{listings}
\usepackage{color}
\usepackage{epsfig,makecell}    
\usepackage{ifpdf}
\ifpdf
  \usepackage{epstopdf}
\fi                                  %%
\usepackage[square,comma,numbers,sort&compress]{natbib}          %%
\usepackage[hyphens]{url}                                         %%
\usepackage[colorlinks]{hyperref}                      %%
\usepackage[usenames,dvipsnames]{xcolor}                          %%
\hypersetup{citecolor=blue,linkcolor=blue}                        %%
\usepackage{amsmath,amsopn,amssymb,amsthm}                        %%
\usepackage{thmtools}
\usepackage[toc,page]{appendix}
\usepackage{subfigure}
\usepackage{endnotes,microtype,xspace,fancyvrb,multirow} %%
\usepackage{booktabs}                                             %%
\usepackage{array,underscore,relsize}                             %%
\usepackage[T1]{fontenc}                                          %%
\usepackage{times}                                                %%
\usepackage{fancyhdr,lastpage}                                    %%
\usepackage{enumitem}                                             %%
\usepackage[export]{adjustbox}                                    %%
\usepackage{breakurl}                                             %%
\usepackage{pifont}                                               %%
\usepackage{tablefootnote}                                        %%
\usepackage[linesnumbered,vlined,ruled,commentsnumbered]{algorithm2e}
\DontPrintSemicolon

\SetCommentSty{mycommfont}

\usepackage{amssymb}
\usepackage[normalem]{ulem}
\usepackage[hang,flushmargin]{footmisc}
\usepackage{textcomp}
\usepackage{graphicx}

\usepackage[compact]{titlesec}
\titleformat*{\section}{\large\bfseries}
\titleformat*{\subsection}{\normalsize\bfseries}
\titleformat*{\subsubsection}{\normalsize\bfseries}
\titlespacing*{\section}{0pt}{*2}{*1}
\titlespacing*{\subsection}{0pt}{*1.5}{*0.8}

\definecolor{mygreen}{rgb}{0,0.6,0}  
\definecolor{mygray}{rgb}{0.5,0.5,0.5}  
\definecolor{mymauve}{rgb}{0.58,0,0.82}

\hypersetup{
  colorlinks,
  linkcolor={red!50!black},
  citecolor={blue!50!black},
  urlcolor={blue!80!black}
}

\newcommand{\squishlist}{
  \begin{list}{$\bullet$}{
    \setlength{\itemsep}{0pt}
    \setlength{\parsep}{3pt}
    \setlength{\topsep}{3pt}
    \setlength{\partopsep}{0pt}
    \setlength{\leftmargin}{1.5em}
    \setlength{\labelwidth}{1em}
    \setlength{\labelsep}{0.5em}
  }
}

\newcommand{\squishend}{
  \end{list}
}

\newcounter{enumctr}
\newcommand{\squishenum}{
  \begin{list}{\arabic{enumctr}.}{
    \usecounter{enumctr}
    \setlength{\itemsep}{0pt}
    \setlength{\parsep}{3pt}
    \setlength{\topsep}{3pt}
    \setlength{\partopsep}{0pt}
    \setlength{\leftmargin}{1.5em}
    \setlength{\labelwidth}{1em}
    \setlength{\labelsep}{0.5em}
  }
}

\usepackage{tikz}

\newcommand{\nospacestitle}[1]{\noindent{\bf #1}}
\newcommand{\stitle}[1]{\vspace{.7ex}\noindent{\bf #1}}
\newcommand{\etitle}[1]{\vspace{0.5ex}\noindent{\em\underline{#1}}}

\newcommand{\sys}[0]{\textsc{DiFache}\xspace}

\newcommand{\us}{{{$\upmu$}s}\xspace}
\newcommand{\x}{{$\times$}\xspace}

\let\oldding\ding%
\renewcommand{\ding}[2][1]{\scalebox{#1}{\oldding{#2}}}%

\usepackage{balance}
\usepackage{upgreek}

\begin{document}

\title{\Large \bf {\sys}: Efficient and Scalable Caching on Disaggregated Memory \\using Decentralized Coherence}

\author{
    Hanze Zhang\;\;\;\;
    Kaiming Wang\;\;\;\;
    Rong Chen\;\;\;\;
    Xingda Wei\;\;\;\;
    Haibo Chen\\[5pt]
 \normalsize{{Institute of Parallel and Distributed Systems, Shanghai Jiao Tong University}} \\ [15pt]
 %\normalsize{\emph{Contacts: {rongchen}@sjtu.edu.cn}}
}

% \author{{Paper \#458}}
% \author{Graduate category}

% \authorinfo{Hanze Zhang\and Ke Cheng\and Rong Chen\and Haibo Chen}
          %  {Institute of Parallel and Distributed Systems, Shanghai Jiao Tong University}
          %  {}

\maketitle
\frenchspacing

\begin{abstract}
  The disaggregated memory (DM) architecture 
  offers high resource elasticity at the cost of data access performance.
  While caching frequently accessed data in compute nodes (CNs) %
  reduces access overhead,
  it requires costly centralized maintenance of cache coherence across CNs. 
  This paper presents {\sys}, an efficient, scalable and coherent CN-side caching framework
  for DM applications.
  Observing that DM applications already serialize conflicting remote data access 
  internally rather than relying on the cache layer,  
  {\sys} introduces \emph{decentralized coherence} that 
  aligns its consistency model with memory nodes instead of CPU caches, thereby
  eliminating the need for centralized management. 
  {\sys} features a decentralized invalidation mechanism %
  to independently invalidate caches on remote CNs and %
  a fine-grained adaptive scheme to cache objects with varying read-write ratios. 
  Evaluations using 54 real-world traces from Twitter show that {\sys} 
  outperforms existing approaches by up to 10.83{\x} {(5.53{\x} on average)}. 
  By integrating {\sys}, the peak throughput of two real-world DM applications 
  increases by 7.94\x and 2.19\x, respectively.
\end{abstract}

\section{Introduction}

The disaggregated memory (DM) architecture, which decouples
CPU and memory into network-attached compute nodes (CNs) and
memory nodes (MNs), is becoming increasingly 
popular~\cite{legoos,polardbsrvless,dmos,ditto,chime}.
Datacenter applications, including commercial products like
PolarDB~\cite{polardbsrvless,polardbmp}, are now specifically designed for DM, 
enabling the independent scaling of compute and memory resources. 
Typically, these applications run clients on CNs that directly access 
data stored in MN memory through network read/write 
operations such as one-sided RDMA~\cite{sherman,ford,smart,chime,ditto,motor}.

\begin{figure}[t]
  \begin{minipage}{1.\linewidth}
    \centering\includegraphics[scale=.42]{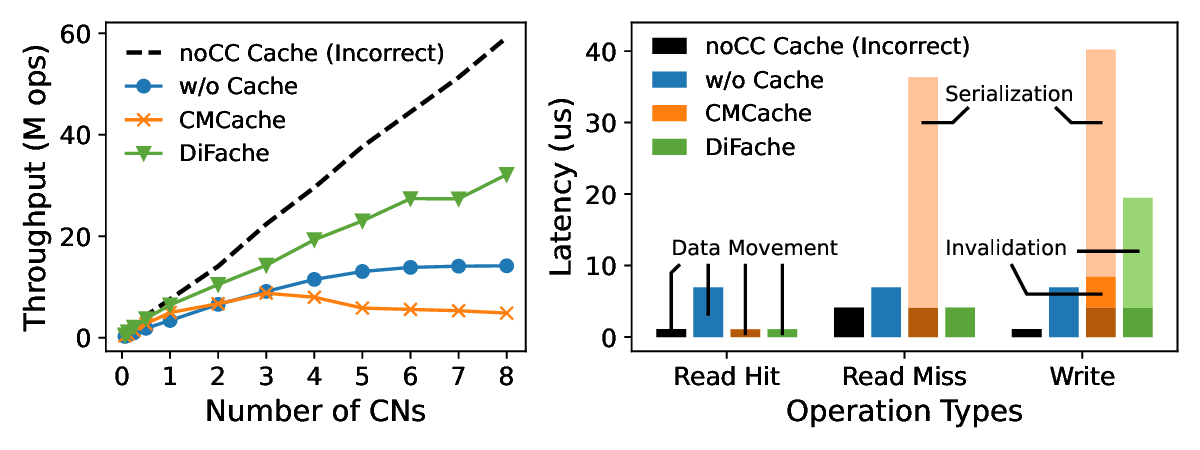}
  \end{minipage} \\[-7pt]
  \begin{minipage}{1.\linewidth}
    \caption{\small{\emph{
      Peak throughput scaling of different caching schemes on DM (left) 
      and median latency breakdown of cache operations (right).
      \textup{\textbf{Workload:}} a real-world Twitter trace (No.\,4) with 93\% reads.
      \textup{\textbf{Testbed:}} 9 CNs and 1 MN, connected with 100\,Gbps RDMA NICs.
    }}}
    \label{fig:intro}
  \end{minipage}  \\[-10pt]
\end{figure}

Although the DM architecture enhances resource elasticity and 
utilization, it significantly degrades the data access 
performance of applications. Specifically, DM applications
must access data over the network, leading to performance
bottlenecks due to network latency and bandwidth 
constraints~\cite{sherman,ford,chime}.
For instance, when executing real-world traces from Twitter~\cite{twittertrace},
the saturation of MN bandwidth often limits throughput scaling
and elevates the latency of moving data from or to MNs
(see ``{w/o Cache}'' in Fig.~\ref{fig:intro}).

Caching remote data in CN memory (referred to as \textbf{CN-side cache})
can effectively reduce data access overhead on DM~\cite{aifm,fastswap,canvas,atlas}.
By converting remote reads and writes to local cache accesses, 
the throughput scales linearly, as it eliminates data movements 
across the network (see ``{noCC Cache}'' in Fig.~\ref{fig:intro}).
However, this approach cannot guarantee that updates to cached data 
on one CN are visible to others---it lacks \textbf{cross-CN cache coherence}.
This limitation can produce incorrect results in 
DM applications that require all clients to see consistent updates, such as 
transactional databases~\cite{ford,motor} and indexes~\cite{sherman,smart,chime}, 
hindering its adoption in practice.

\begin{figure}[t]
  \begin{minipage}{1.\linewidth}
    \centering\includegraphics[scale=.59]{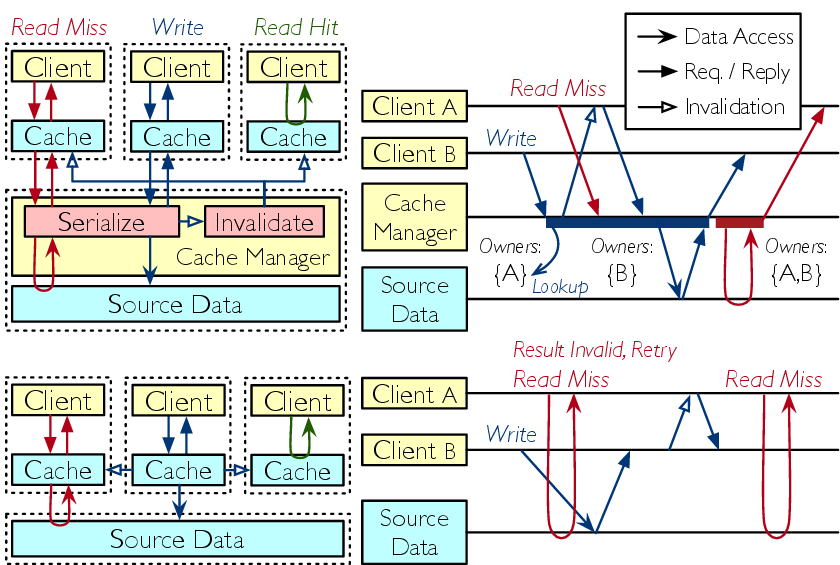}
  \end{minipage} \\[-2pt]
  \begin{minipage}{1.\linewidth}
    \caption{\small{\emph{The architecture and workflow of centralized (top)
    and decentralized (bottom) designs of cache coherence.
    }}}
    \label{fig:bsl}
  \end{minipage}  \\[-10pt]
\end{figure}

An intuitive approach to maintaining cache coherence across distributed nodes 
is to implement CPU cache designs in software~\cite{ivy,gam,polardbsrvless}.
When a write occurs on one node, it invalidates cached data on other nodes, 
as illustrated in Fig.~\ref{fig:bsl} (top).
Clients attempting to read from invalidated caches experience 
read misses and must retrieve updated data from remote nodes.
A centralized manager handles all writes and read misses 
to ensure linearizability of data accesses. %
This manager \emph{serializes} data movements between CN-side caches and MNs, 
as well as \emph{invalidates} outdated cached data.
To limit invalidation costs as the number of CNs increases, 
the manager keeps track of nodes with caches (i.e., \emph{owners}) 
and invalidates only those nodes during writes.

Unfortunately, when applied to DM environments, this approach provides
restricted benefits from caching and suffers from performance degradation 
with a large number of CNs (see ``{CMCache}'' in Fig.~\ref{fig:intro}). 
This happens because all cache operations are handled by a centralized manager, 
which becomes %
a significant bottleneck as the operation rate increases. 
Even when the manager saturates all 16 cores on a dedicated CN, 
the serialization cost---primarily from the queuing delay of 
cache operations---still accounts for nearly 90\% of the median 
operation latency when 8 CNs are involved.

\begin{figure*}[!ht]
  \begin{minipage}{1.\linewidth}
    \centering\includegraphics[scale=.4]{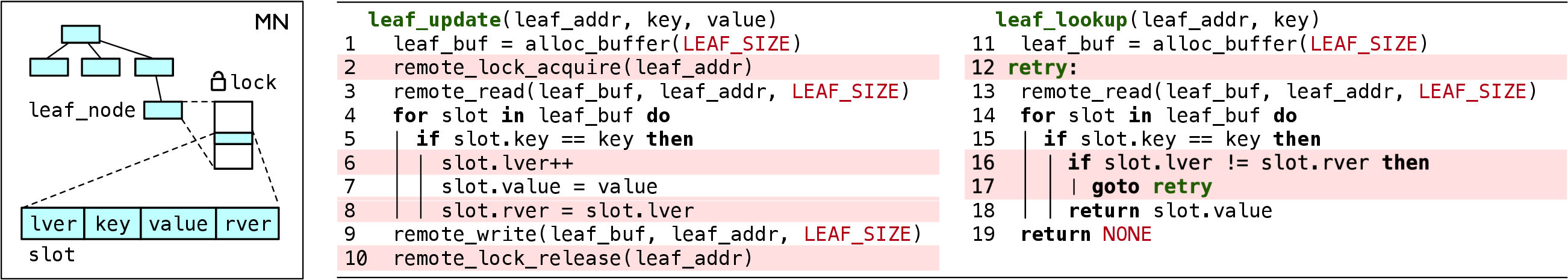}
  \end{minipage} \\[-2pt]
  \begin{minipage}{1.\linewidth}
    \caption{\small{\emph{%
	Simplified code snippets of a DM application 
	demonstrating update and lookup operations for leaf nodes in a 
  DM-based tree index (Sherman~\cite{sherman}). 
	The highlighted codes with a red background serialize remote accesses.
    }}}
    \label{fig:app}
  \end{minipage}  \\[-10pt]
\end{figure*}

\stitle{Opportunity.}
{DM applications have weak consistency requirements for remote data access}
because they recognize that their reads and writes occur over {the network
rather than the local memory bus}. As a result, these applications always 
serialize conflicting remote data access internally rather than relying on 
the cache layer~\cite{clover,polardbmp,ford,rtx,motor,sherman,smart,chime,rolex,fusee,ditto,race}. 
For example, in Fig.~\ref{fig:app}, when accessing a DM-based tree index,
concurrent remote writes are prohibited by locking, and remote reads  
are retried if the retrieved data is inconsistent. %

\stitle{Key insight.}
\emph{The consistency model of CN-side caches for DM applications 
can align with MNs rather than CPU caches}. 
This allows the clients to independently maintain cross-CN cache coherence 
in a decentralized design, eliminating the need for a centralized manager 
to serialize cache accesses.

\stitle{Our approach.}
Based on this insight, we present {\sys}, 
an efficient, scalable and coherent CN-side caching framework
for DM applications.
It translates remote operations in the application
to its corresponding cache operations, accelerating remote data
accesses without application redesigns. The key idea of {\sys} is
\emph{decentralized coherence}, where cache coherence is maintained
independently by clients instead of a centralized manager,
as shown in Fig.~\ref{fig:bsl} (bottom). 
For reads, clients fetch data from the cache on hits or directly 
from MNs on misses. 
For writes, clients flush modified data to MNs first, and then 
invalidate caches on other nodes. %
Compared to centralized approaches, decentralized coherence
eliminates serialization costs
and distributes invalidation tasks across individual clients.
Consequently, the performance of decentralized coherence is now 
largely dominated by invalidation costs---a crucial factor for achieving 
efficient and scalable caching on DM (see ``{DiFache}'' in Fig.~\ref{fig:intro}). 
{\sys} is novel in two key ways to address this challenge. 

\stitle{Decentralized invalidation.}
To maintain cache coherence independently, clients must update cache metadata 
in the remote memory of other CNs (owners) while bypassing their CPUs.
However, since cached objects are allocated and evicted
dynamically, their addresses %
cannot be predetermined. %
To tackle this, each CN uses a hopscotch hash table to index cached objects, 
enabling address lookup with a single remote operation.
Furthermore, without a centralized manager to track owners, 
invalidations must be broadcast to all CNs, which can cause 
substantial network traffic when many CNs are involved. 
To enhance scalability, when the number of CNs exceeds a threshold,
clients record owners in remote bitmaps during read misses 
to avoid unnecessary invalidations.

\stitle{Fine-grained adaptive caching.}
Enabling CN-side cache significantly improves read performance
but introduces inevitable cache coherence costs---primarily invalidation costs 
in {\sys}---for writes, as shown in Fig.~\ref{fig:intro}(right). 
Therefore, when writes dominate the workload, overall performance gains 
decline rapidly and can even become negative. %
Worse still, we discovered that objects in real-world workloads often 
have varying read-write ratios
(see Fig.~\ref{fig:select}) and short access periods,
making a simple binary choice of enabling or disabling CN-side cache 
across the entire system suboptimal.
To address this, {\sys} models the profits of caching for each object 
based on its hit rates and read ratios, which are collected using
lightweight statistics measured in nanoseconds. %
Based on these periodically calculated profits, clients can dynamically 
enable or disable caching for individual objects,  
adapting to varying workloads without manual involvement. %

\stitle{Evaluation.}
We evaluated {\sys} using 54 real-world in-memory caching traces from 
Twitter~\cite{twittertrace} and two real-world DM applications: 
a database index (Sherman~\cite{sherman}) and a transaction 
engine (FORD~\cite{ford}) with representative workloads.
Our experimental results show that {\sys} outperforms no caching
and centralized caching by up to 8.16\x and 10.83\x 
(1.85\x and 5.53\x on average) when running
Twitter traces. When integrated with Sherman and FORD, {\sys} further
improves their peak throughput by up to 7.94\x and 2.19\x.

\stitle{Contributions.}
We summarize our contributions as follows:

\squishlist
\vspace{-.4mm}
\item A new {decentralized} design that enables coherent CN-side caching 
  in DM applications.
\vspace{-.4mm}
\item A {decentralized invalidation} mechanism for %
  CPU-efficient and scalable cache invalidation on remote CNs.
\vspace{-.4mm}
\item A {fine-grained adaptive caching} scheme that 
  dynamically identifies objects whose caching overhead outweighs
  benefits and disables caching for them.
\vspace{-.4mm}
\item An evaluation using real-world traces and DM applications
  that shows the advantage and efficacy of {\sys}.
\squishend

\section{Background and Motivation}\label{sec:bg}

\subsection{DM Architecture and Applications}\label{sec:bg:app}

In the disaggregated memory (DM) architecture, CPU and memory resources are
separated into compute nodes (CNs) and memory nodes (MNs), connected by
a unified high-speed network~\cite{dmnetwork,dmos,legoos}.\footnote{\footnotesize{Without 
loss of generality, this paper considers an RDMA network by default
like prior DM applications~\cite{sherman,smart,ditto,chime,ford,rolex}.
We discuss CXL-based DM in Appendix A (see our supplementary materials).}}
Applications run client threads on CNs, which have massive CPU 
cores but limited DRAM (a few GB~\cite{fusee,smart}), 
and store %
their data on MNs, which possess sufficient memory but negligible 
computing power. This separation allows for more flexible resource 
allocation, enhancing resource elasticity and utilization.
To embrace these advantages, applications such as OLTP 
databases~\cite{ford,rtx,motor,polardbsrvless}, 
key-value stores~\cite{clover,fusee,ditto,race}, and tree-based
indexes~\cite{sherman,smart,rolex,chime} have been adapted to DM environments.

Fig.~\ref{fig:app} illustrates the simplified data structures and 
operations of a DM tree-based index (Sherman~\cite{sherman}) as an example.
DM applications typically access MN data at \emph{object} granularity,
meaning they read or write an entire object (leaf nodes in this example)
with a single remote operation (Lines 3, 9, and 13).
Finer-grained accesses are performed locally (Lines 4--8 and 14--18).
Since remote accesses are not 
linearizable~\cite{herlihy1990linearizability}---meaning 
a read or write may start before another write on the same object 
finishes---DM applications 
always serialize remote accesses internally.
Concurrent writes on the same object can corrupt data and
are strictly prevented, typically using locks (Lines 2 and 
10)~\cite{clover,polardbmp,ford,rtx,motor,sherman,smart,chime,rolex} 
or out-of-place updates~\cite{fusee,ditto,race}.
For reads, some applications pessimistically serialize them with writes 
using reader-writer locks~\cite{clover,polardbmp} or 
read-copy-update mechanisms~\cite{fusee,ditto,race}.
Others, like this example, allow lock-free reads and validate the results 
using per-object versions 
(Lines 6, 8, and 16--17)~\cite{ford,rtx,motor,sherman,smart,chime,rolex}.

While achieving higher resource elasticity and utilization, DM applications
face significant challenges related to data access performance. 
Despite using CPU-bypassing techniques (e.g., RDMA and CXL~\cite{cxl}) 
to directly access data in MN memory, they still experience significantly
higher data access latency (hundreds or thousands of nanoseconds) 
and lower bandwidth (tens of GB/s) compared to local memory 
access~\cite{rdmastudy,hydrarpc,cxltxn}. 
Consequently, remote data accesses often become the performance bottleneck 
in DM applications.

\subsection{CN-side Caching for DM Applications}\label{sec:bg:cache}

An intuitive approach to mitigate the overhead of accessing remote memory
is to cache frequently accessed data in CN 
memory~\cite{aifm,kona,atlas,infiniswap,fastswap,canvas}.
Remote reads and writes are redirected to the cached data when possible,
leading to lower access latency and reduced MN bandwidth consumption, 
thereby enhancing application performance.
When the CN-side cache becomes full, cached data is swapped with other
data in remote memory based on specific cache policies, such as 
LRU~\cite{fastswap,canvas}.
The CN-side cache can be managed at the operating system level, swapping
cached data in \emph{page} granularity to remain oblivious to application 
semantics~\cite{infiniswap,fastswap,canvas}.
Another approach is application-aware, managing the cache in \emph{object}
granularity to reduce network consumption during 
swapping~\cite{aifm,kona,atlas}.
In practice, existing CN-side caching approaches only allow clients 
from a single CN to share cached data~\cite{atlas,aifm,kona,infiniswap,
fastswap,canvas}.

\begin{figure}[t]
  \begin{minipage}{1.\linewidth}
    \centering\includegraphics[scale=.33]{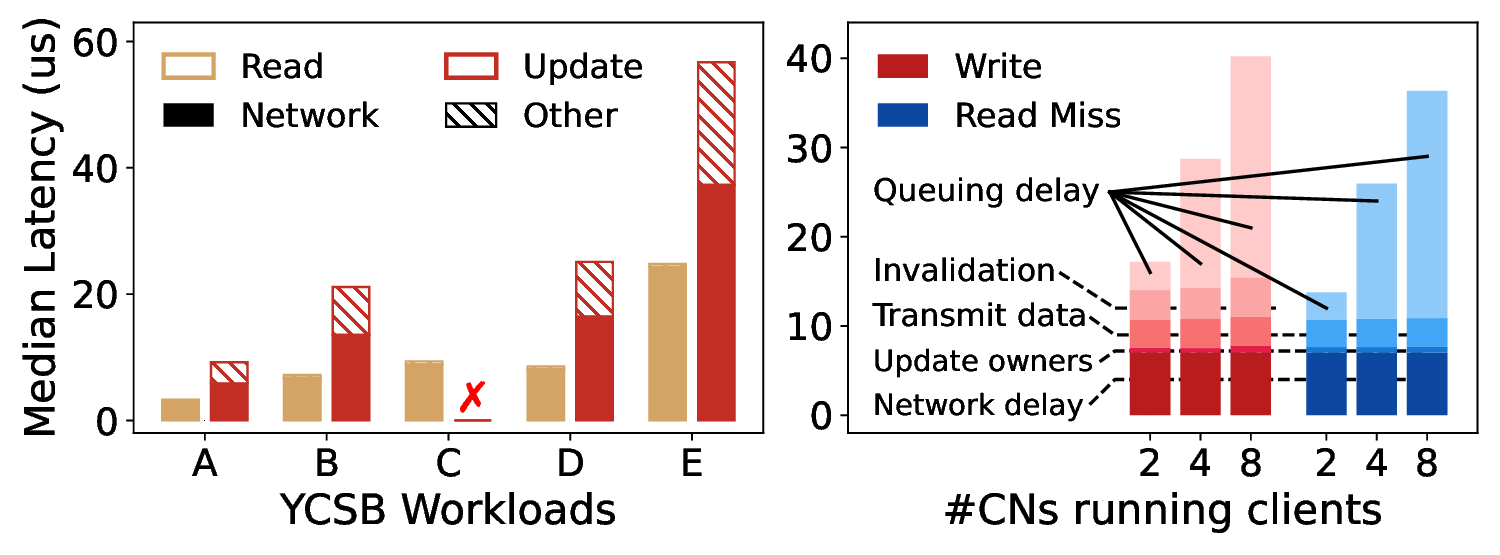}
  \end{minipage} \\[-7pt]
  \begin{minipage}{1.\linewidth}
    \caption{\small{\emph{Median latency breakdown of read and 
    update operations 
    in Sherman running YCSB workloads (left), and that of 
    writes and read misses in CMCache running the No.\,4 Twitter trace
     (right).
    YCSB C has no update latency because it is read-only.
    }}}
    \label{fig:motiv}
  \end{minipage}  \\[-10pt]
\end{figure}

\stitle{Problem: Lacking cross-CN cache coherence.}
DM applications often scale out to multiple CNs to allocate
sufficient computing resources, requiring all CNs to see consistent
data updates~\cite{sherman,smart,chime,ditto,polardbsrvless}.
This necessitates \emph{cross-CN cache coherence}, which
existing approaches of CN-side caching lack support for.
Consequently, most DM applications restrict caching to small metadata 
that can be easily identified as outdated~\cite{sherman,chime,ford,ditto}. 
For example, the tree-based index caches the address of leaf nodes 
and identifies outdated addresses by validating the contents of nodes 
retrieved using those addresses~\cite{sherman,chime}. 
However, since large data like leaf nodes are still accessed via the 
network, network delays continue to dominate the operation latency of 
applications.
As shown in Fig.~\ref{fig:motiv} (left), when running YCSB workloads
with 128 clients, network delays account for at least 87.2\% of read 
latency and 63.3\% of update latency of Sherman~\cite{sherman}.

\subsection{Distributed Cache Coherence}\label{sec:bg:dcc}

Traditionally, cache coherence between distributed nodes is maintained
by simulating hardware cache functionalities using a centralized software 
manager~\cite{ivy,gam,mind}. This manager \emph{serializes}
concurrent reads and writes, similar to a CPU bus, and \emph{invalidates} cached
data on distributed nodes upon writes, like a cache directory~\cite{cachedir}.
This approach is also adopted by a DM application,
PolarDB-MP~\cite{polardbmp}, for maintaining cross-CN cache coherence.
As shown in Fig.~\ref{fig:bsl} (top),
when the application issues remote read operations, if the target object
is cached (\emph{Read Hit}), it copies the object directly from the cache.
Otherwise (\emph{Read Miss}), an RPC request is sent to the manager
for the latest data, which then marks the requesting CN 
as an \emph{owner} and provides the data. 
All remote write operations require an RPC to the manager, which invalidates 
caches on existing owners, sets the writer as the sole owner, and 
updates the source data. To prevent racing accesses to owners or the source 
data, the manager serializes RPCs for read misses and writes to the same object.

\stitle{Problem: Inefficient centralized management.}
In DM environments with a highly elastic number of clients, the
centralized manager often becomes overwhelmed handling RPCs for read 
misses and writes, leading to significant performance bottlenecks.
Typically, the manager is placed on MNs to co-locate with the source 
data~\cite{polardbsrvless,polardbmp}, leading to hundreds of
microseconds of cache operation latency on our testbed due 
to MNs' limited processing power. 
Even when the manager's processing power is enhanced by running 
it on a dedicated CN with all 16 cores, each RPC still takes tens of
microseconds to complete, causing
throughput degradation with an increasing number of clients,
as shown in Fig.~\ref{fig:intro}.

To identify the cause, we break down the median latency of cache 
operations in Fig.~\ref{fig:intro}.
As shown in Fig.\ref{fig:motiv} (right), the latency is dominated
by serialization costs,
i.e., \emph{network delay} of sending the RPC and receiving its 
reply and the \emph{queuing delay} of RPCs that have arrived at 
the manager but cannot be handled immediately. 
While the network delay is stable, the queuing delay
is proportional to the number of clients, which significantly
hampers the scalability of caching.

\section{Approach and Overview}\label{sec:ov}

\nospacestitle{Opportunity: Weak consistency requirements.}
DM applications access MN data at object granularity using explicit
network operations. Since network operations are not linearizable, 
these applications commonly serialize remote accesses 
to the same object internally~\cite{clover,polardbmp,ford,rtx,motor,
sherman,smart,chime,rolex,fusee,ditto,race}. To accelerate these
accesses with weak consistency requirements, CN-side caches do 
not need to guarantee the linearizability of cache operations.
This allows clients to exchange data with MNs and invalidate caches 
on other CNs independently, without relying on a centralized manager
to serialize these operations, avoiding its inefficiencies. 

DM applications tolerate non-linearizable remote accesses by
serializing them with techniques introduced in \S\ref{sec:bg:app},
which ensure the correctness of decentralized cache operations,
as shown in Fig.~\ref{fig:bsl} (bottom). Because writes to the
same object are strictly ordered, only one client can flush the 
object to MNs and invalidate it on other CNs at a time,
preventing interference. In pessimistic 
designs~\cite{clover,polardbmp,fusee,ditto,race},
reads are isolated from writes, ensuring that they
always see the latest object. However,
in optimistic designs~\cite{ford,rtx,motor,sherman,smart,chime,rolex},
reads may interleave with writes. If the client retrieves a partially
written object during a read miss, the application can detect the
inconsistency (e.g., using versions) and retry until a consistent 
object is obtained. These retries hit the cache until it is
invalidated by the write, triggering a read miss that corrects
the results. Since writes invalidate other CNs only after flushing 
updates to MNs, all CNs are guaranteed to retrieve the updated object
after invalidation.

\begin{figure}[t]
  \vspace{1mm}
  \begin{minipage}{1.\linewidth}
    \centering\includegraphics[scale=.53]{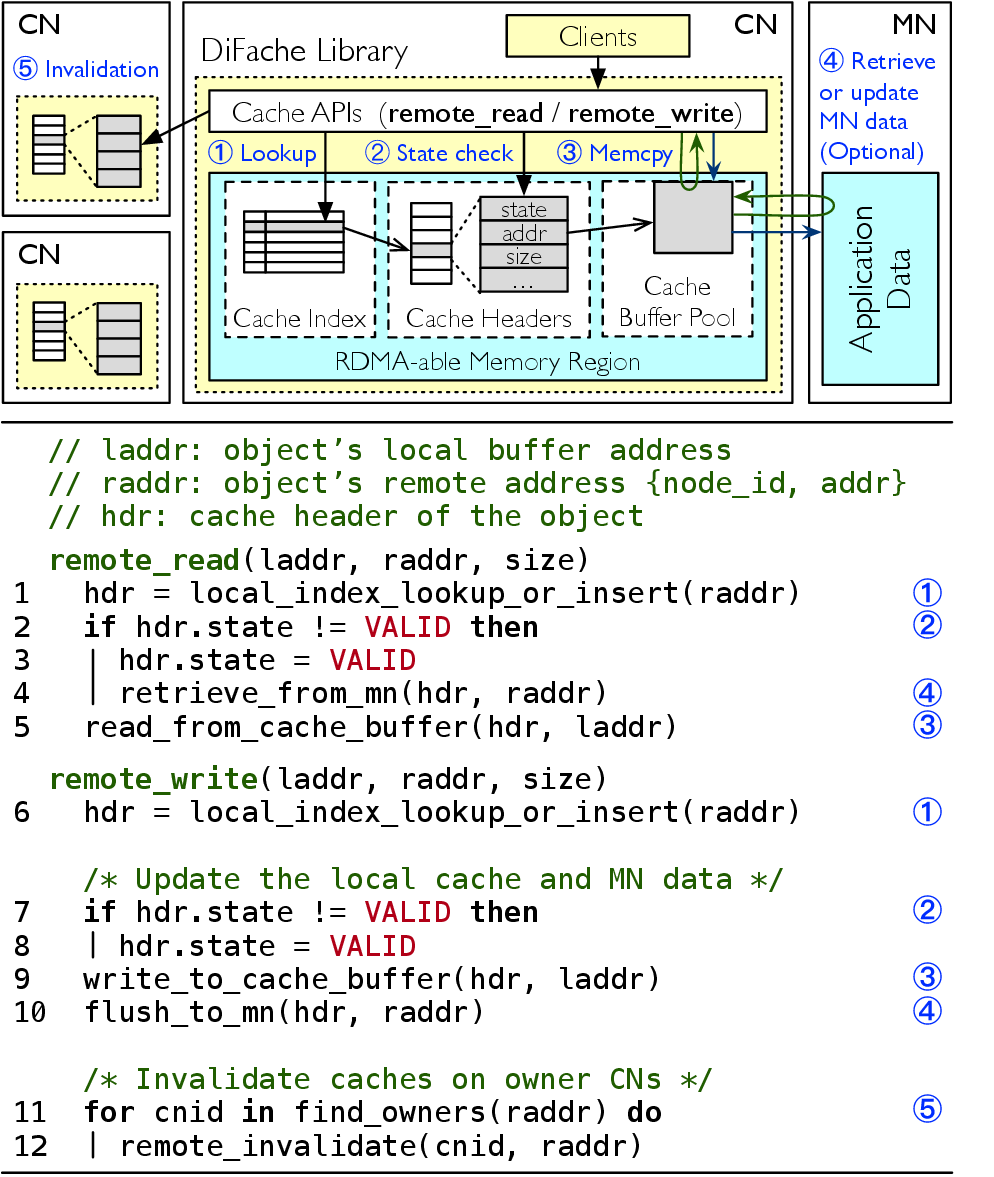}
  \end{minipage} \\[-2pt]
  \begin{minipage}{1.\linewidth}
    \caption{\small{\emph{The architecture (top) and APIs (bottom) of {\textup{\sys}}.
	}}}
    \label{fig:arch}
  \end{minipage}  \\[-15pt]
\end{figure}

\stitle{Our approach.}
Leveraging this opportunity, we propose \emph{decentralized coherence} 
to thoroughly eliminate the inefficiencies of centralized management 
for DM applications, as shown in Fig.~\ref{fig:bsl} (bottom).
Correctness is ensured by aligning the consistency model of CN-side 
caches with that of MNs---exchanging data with MNs and maintaining 
cross-CN cache coherence at the object level. 
For reads, clients retrieve the object from the cache on hits or directly 
from MNs on misses. For writes, clients flush the updated object to MNs,
and then invalidate the cached object on other CNs.
Decentralized coherence distributes invalidation tasks across individual
clients, thereby avoiding the significant network and queuing delays 
associated with centralized managers.

\stitle{Overview of {\sys}.}
{\sys} is a caching framework that implements decentralized coherence
for efficient and scalable CN-side caching in DM applications. 
It caches application data at the \emph{object} level and identifies
objects by their remote addresses, which consists of 
the MN's ID and the object's address on that MN.
{\sys} provides APIs consistent with remote operations, 
enabling applications to integrate it seamlessly 
by switching their implementation of remote reads and writes.
{\sys} operates independently of DM applications' memory allocation, 
determining object addresses and sizes directly from remote operation parameters.

\etitle{Architecture.} As shown in Fig.~\ref{fig:arch} (top),
each CN uses a cache index to map objects' remote addresses
to cache headers, which records metadata such as cache state
(\emph{state}), local buffer address (\emph{addr})
storing the cached object, and object size (\emph{size}). 
Buffers for cached objects are allocated from a 
\emph{cache buffer pool}. 
The cache index, headers, and buffer pool are located
in an RDMA-accessible memory region. 

\etitle{Workflow.} As shown in Fig.~\ref{fig:arch} (bottom),
during read operations, the client first looks up the local 
cache index for the object's cache header (Line 1). If the 
object lacks a recorded header, a new one is allocated and 
added to the index. If the header state
is valid, the client copies the object directly from 
the cache buffer (Line 5). Otherwise, it sets the state to valid,
retrieves the object from MNs, and then performs the copy 
(Lines 3--5). For write operations, after 
locating the cache header (Line 6), the client flushes updates
to both the cache buffer and MNs (Lines 7--10) and
invalidates the cached object on owner CNs (Lines 11--12).

\stitle{Challenge.}
After eliminating serialization costs from the centralized manager,
{\sys}'s performance is dominated by invalidation costs,
which face new challenges in two ways.

First, for efficient invalidation, clients must 
update cache states on remote CNs while bypassing their CPUs.
However, since cached headers are allocated and evicted dynamically, 
the addresses of cache states cannot be predetermined. 
Moreover, %
without tracking which CNs have valid caches (owners),
clients must invalidate all other CNs, generating significant 
network traffic when there are many CNs. 
{\sys} proposes a \emph{decentralized invalidation} mechanism that 
locates remote cache states and tracks owners efficiently
(\S\ref{sec:design}).

Second, 
as the proportion of writes in the workload increases, the performance 
benefits of caching can decline rapidly and even become negative 
due to higher invalidation costs and lower hit rates. 
Worse yet, objects in real-world workloads often have varying read-write 
ratios and short access periods. %
Thus, a simple binary choice of enabling or disabling caching
across the entire system is suboptimal.
{\sys} proposes a \emph{fine-grained adaptive caching} scheme that 
applies the suitable cache mode for each individual object based on 
its dynamically estimated caching profits (\S\ref{sec:selective}).

\section{Decentralized Invalidation}\label{sec:design}

To efficiently obtain remote addresses of cache states on other CNs (owners) 
for decentralized invalidation, 
{\sys} introduces a hopscotch-based cache index for remote lookups %
and combines broadcasting with owner sets to track owners. %

\subsection{Hopscotch Cache Index}\label{sec:design:index}

To select an appropriate data structure for the cache index,
we identify two key attributes based on its role in {\sys}:

\squishlist
\vspace{-.5mm}
\item \textbf{Frequent remote lookups (A\#1)}: 
  Clients look up the cache index on each remote owner during writes,
  rendering its remote lookup efficiency crucial.
  
\vspace{-.5mm}
\item \textbf{Scarce and local insertions (A\#2)}: 
  Once inserted, cache headers remain in the cache index until they are 
  evicted, making insertion a rare event that occurs only locally.
  This allows for more complex insertion logic.
\squishend

We find that \emph{hopscotch hashing}~\cite{hopscotch} 
is well-suited for the cache index, as it ensures that a key-value 
pair can always be inserted and found within a specific region 
starting from the hashed key.
This feature allows remote clients to look up the index with 
\emph{a single} remote read operation that retrieves the entire 
region, which satisfies A\#1. Although the insertion process is 
complex due to the need to move the key-value pair into the region, 
this complexity is tolerable because of A\#2.

\begin{figure}[t]
  \begin{minipage}{1.\linewidth}
    \centering\includegraphics[scale=.55]{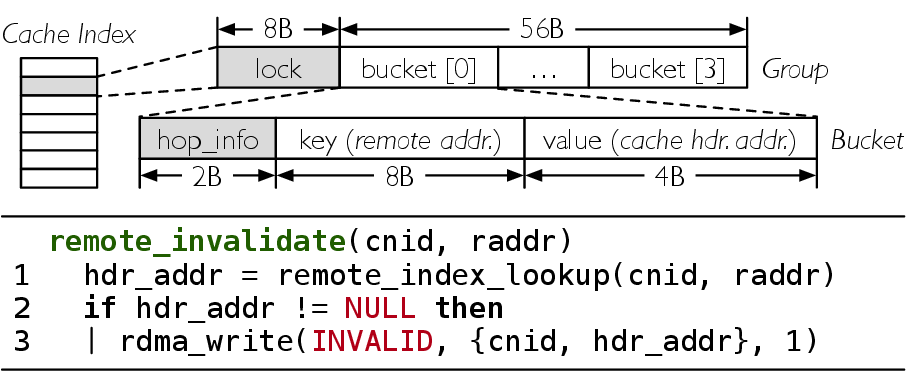}
  \end{minipage} \\[-2pt]
  \begin{minipage}{1.\linewidth}
    \caption{\small{\emph{The cache index structure based on hopscotch 
    hashing (top) and the pseudo-code of remote invalidation (bottom).
    }}}
    \label{fig:hash}
  \end{minipage}  \\[-10pt]
\end{figure}

\stitle{Data structure.}
Fig.~\ref{fig:hash} (top) illustrates the structure of the cache index.
It consists of an array of \emph{buckets}, each
storing a key-value pair. A hash function maps a given key
to a specific bucket, known as the \emph{home bucket}, in the array. 
The home bucket of a key has the same offset in all CNs' memory regions, 
allowing clients to directly calculate its address on remote CNs.
Hopscotch hashing ensures that any key-value pair can be found within
the $H$ consecutive buckets starting from its home bucket, known as 
its \emph{neighborhood}.

\etitle{Key and value types.}
The cache index uses the cached object's remote 
address as the key and the cache header's local address as the value.
For higher space efficiency, the hash table stores the
cache header's offset in {\sys}'s RDMA-accessible memory region instead of
the full virtual address. Since CNs typically have at most
a few GBs of memory, 4 bytes are sufficient for storing each value.

\etitle{Bucket structure.}
Each bucket stores a key-value pair
and a bitmap called \texttt{hop_info} that tracks the occupancy
of its neighborhood. If the $i$th bit of \texttt{hop_info} is set to 1,
it indicates that the current bucket is the home bucket for the 
key-value pair in the $i$th bucket of the neighborhood. We assign two 
bytes to \texttt{hop_info}, allowing for a neighborhood size ($H$) of 
16, which can support a 99\% load factor~\cite{chime}. 

\etitle{Grouped buckets.}
Adjacent buckets in the cache index are merged into 64-byte
\emph{groups} aligned with CPU cache lines, ensuring the 
consistency of their contents when being retrieved by remote 
clients~\cite{farm}.
Each group contains four buckets and a mutex lock for serializing
updates to the group. 

\stitle{Lock-based insertion.}
Insertions to the cache index follow the hopscotch hashing 
algorithm~\cite{hopscotch}.
To insert a key-value pair, the client searches the array of buckets 
for an empty one, starting from the home bucket determined by 
hashing the key. Groups of the searched buckets
are locked in sequence to serialize updates.
All clients lock groups in ascending order to avoid deadlocks.
If a bucket contains the same key as the one being inserted, 
indicating a duplicated insertion, the client directly returns
the value in that bucket and cancels the ongoing insertion. 
Once an empty bucket is found,
the client attempts to move it into the key-value pair's neighborhood
by swapping its content with preceding buckets,
ensuring no key-value pair moves out of its neighborhood. 
The \texttt{hop_info} in swapped buckets is updated accordingly.
If the empty bucket can be moved into the neighborhood after 
several swaps,
the client fills it with the key-value pair, updates the 
\texttt{hop_info} in the home bucket, and releases all locks.
Otherwise, the insertion fails, indicating that the index is full.

\stitle{Lock-free lookup.}
When a local lookup fails, the client attempts to insert a new cache
header. Since duplicated insertions are cancelled, this design allows
{\sys} to tolerate false negatives in local lookups, enabling a
lock-free design that minimizes latency. During local lookups,
the client searches for a bucket with a matching key in the neighborhood
and returns its value, using \texttt{hop_info} to speed up the process.
If no matching bucket is found, the lookup fails.
To ensure the validity of values in matching buckets, insertions always 
update the value before the key when filling an empty bucket, 
and update the key before the value when setting a bucket to empty,
eliminating false positives. 

Remote lookups are also lock-free by utilizing the grouped layout of buckets.
During remote lookups, the client retrieves all groups containing
the neighborhood with a single remote read operation. If any retrieved group 
is locked, the client retries the read to ensure the groups are not 
being updated. The client then searches for a matching bucket
in the neighborhood, similar to the local case.
If a remote lookup fails during remote invalidations, the client does not
invalidate the remote CN because it is not an owner
(see Fig.~\ref{fig:hash} (bottom)).

\stitle{Eviction.}
{\sys} attempts to evict cache headers from the index when an insertion 
fails. During evictions, it examines each bucket in the neighborhood to
find a victim for eviction following specific policies (see details in
\S\ref{sec:impl}), locking all groups containing the examined buckets. 
To evict the victim, it first clears the old key to prevent lookups 
for the new key from reading the old value, then fills the new value 
and key into the bucket in sequence. The cache buffers of the victims are
reclaimed after the eviction.

\subsection{Owner Tracking}\label{sec:design:set}

Inspired by the design principles of CPU cache, {\sys} uses different
approaches to track owners based on the number of CNs.
When the number of CNs is small, it adopts a \emph{broadcast} approach
similar to snoopy caching~\cite{snooping}, where all CNs are considered 
owners. This eliminates the overhead of recording owners during read misses
and checking owners during writes. With a larger number 
of CNs, similar to directory-based caching~\cite{cachedir}, 
it employs \emph{owner sets} to record owners, reducing
unnecessary invalidations during writes. 
Details of how {\sys} dynamically switches between both approaches 
when CN numbers change are described in \S\ref{sec:impl}.

\stitle{Data structure of owner sets.}
To enable decentralized invalidation, owner sets must support 
efficient CPU-bypassing updates. Therefore, {\sys} uses atomic 
bitmaps for owner sets, requiring only one 
compare-and-swap operation to insert owners or clear the set, unlike 
arrays or trees that need multiple remote operations.
To address capacity limitations, {\sys} uses the \texttt{CNID\%N} bit to 
indicate an owner's presence in the set, where \texttt{N} is the
number of bits in the bitmap.
This design may lead to false positives, where a non-owner 
is mistakenly regarded as an owner, but this is acceptable because 
invalidating a non-owner does not affect correctness.
{\sys} uses 64-bit bitmaps because it is the maximum size allowed for 
atomic RDMA operations. 

\stitle{Owner set management.}
Each object has a corresponding owner set on the MN storing it. 
To remain transparent to applications, {\sys} indexes owner sets
using a RACE hash table~\cite{race} on each MN, which supports
atomic remote insertion without involving MN CPUs.
Clients insert an empty owner set into the hash table when
allocating the cache header for an object. RACE hashing ensures 
that duplicated insertions fail and return the existing owner set.

\stitle{Owner set accesses.}
{\sys} accesses owner sets on two occasions.
During read misses, the client inserts its CN ID into the bitmap
before setting the cache state to valid (Line 3 in Fig.~\ref{fig:arch}).
This order ensures that any CN with a valid cache state is in the 
owner set and can be correctly invalidated.
During writes, the client collects owners from the bitmap
and invalidates them after flushing modifications to MNs
(Line 11 in Fig.~\ref{fig:arch}). When collecting owners,
it clears the bitmap and inserts its CN ID atomically
to become the only new owner.

Both types of accesses are supported by a remote compare-and-swap 
(CAS) operation, which updates the remote bitmap only if its original 
value matches an expected value. This operation returns the original
value regardless of whether the bitmap is updated.
In the beginning, the client looks up the owner set's remote address 
from the RACE hash and gets its current value.
During read misses, the client sets the bit corresponding to the 
CN ID to 1 and attempts to swap the remote value with the updated 
value using a CAS operation. If it fails, indicating concurrent 
updates, the client retries using the latest owner set value 
returned by the CAS operation until succeeding. 
During writes, the client tries swapping the remote value with a 
bitmap containing only its CN ID until succeeding. 
It then collects owners by skimming through the IDs of all living
CNs to check whether they belong to the returned bitmap.

\section{Fine-grained Adaptive Caching}
\label{sec:selective}

\begin{figure}[t]
  \begin{minipage}{1.\linewidth}
    \centering\includegraphics[scale=.42]{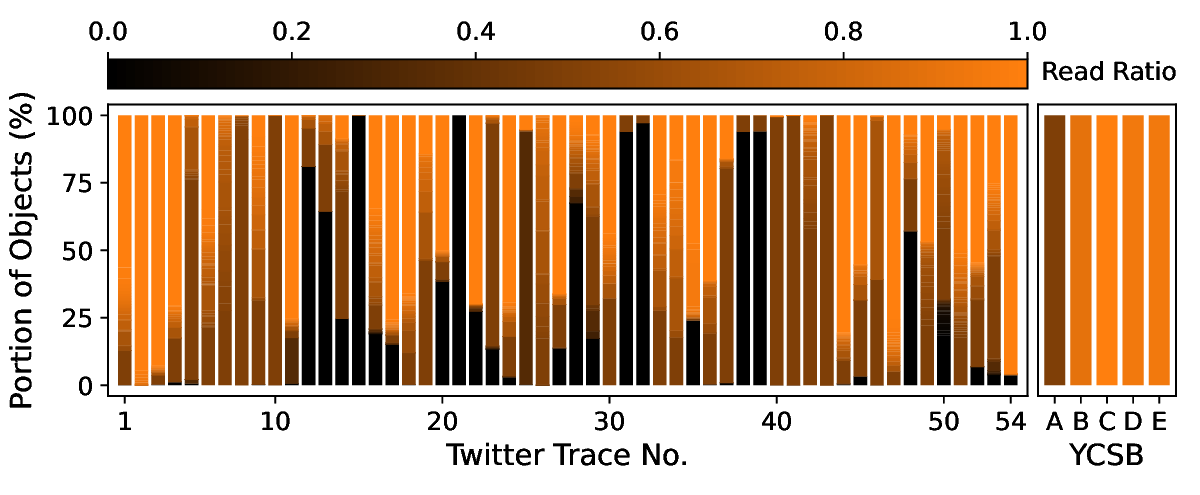}
  \end{minipage} \\[-7pt]
  \begin{minipage}{1.\linewidth}
    \caption{\small{\emph{The read ratio distribution across objects in
    Twitter traces and YCSB workloads.
    }}}
    \label{fig:select}
  \end{minipage}  \\[-10pt]
\end{figure}

The features discovered in real-world workloads motivate 
fine-grained adaptive caching in {\sys}, which
dynamically selects and switches cache modes for each object.

\subsection{Analysis of Real-world Traces}\label{sec:select:analysis}

Our analysis of 54 real-world traces from Twitter's in-memory caching 
clusters~\cite{twittertrace,twittertracegithub} 
revealed two important features for data caching that are often 
overlooked in synthetic workloads of widely-used benchmarks like YCSB~\cite{ycsb}.

\stitle{Observation 1: Objects have varying read-write ratios.}
Objects within the same real-world trace exhibit diverse read-write ratios, 
and this distribution varies significantly across different traces, 
as shown in Fig.~\ref{fig:select}.
This contrasts sharply with the synthetic workloads, such as YCSB~\cite{ycsb}, 
which applies a uniform read-write ratio to all objects.
These findings motivate a fine-grained approach that adaptively enables 
or disables caching for each individual object, 
rather than applying a uniform policy across all objects.
The major challenge is precisely estimating caching profits to select 
the appropriate cache mode for each object (\S\ref{sec:select:thresh}).

\stitle{Observation 2: Objects have short access periods.}
In real-world traces, objects are typically accessed intensively during a short time period,
making it impractical to predetermine a suitable cache mode. %
For example, our analysis of the first million operations in
the No.\,22 Twitter trace showed that nearly 90\% of objects were 
accessed within just 5\% of the total trace duration.
This pattern differs significantly from synthetic workloads, 
like YCSB~\cite{ycsb}, 
where all objects are continuously accessed all the time.
Such short access periods require dynamic cache mode selection and switching 
based on real-time statistics. However, this poses a challenge 
in enabling or disabling caching atomically for an object across all CNs 
(\S\ref{sec:select:switch}).

\subsection{Modeling Caching Profits}
\label{sec:select:thresh}

The caching profits can be described as the difference in average 
data access latencies when caching is enabled versus disabled.
However, these latencies are heavily influenced by dynamically
changing hit rates and read ratios.
Since caching cannot be simultaneously enabled and disabled,
directly comparing latencies from the same time slice is
hard, rendering the average latency a suboptimal metric 
in terms of precision.

We observe that cache operation latencies for each event
(read hit, read miss, and write) remain relatively stable despite
changes in their occurrence ratios. Therefore, {\sys}
breaks down the average latency by summing the product of each
event's ratio ($R$) and its operation latency ($T$). The caching 
profit ($P$) can be modeled as follows, where $rb$ and $wb$ 
indicate reads and writes bypassing the cache, while $rhit$, 
$rmiss$, and $wcached$ indicate cache accesses:

\vspace{-5mm}
\begin{equation*}
  \small
  \begin{split}
    P =\: & R_{rhit}\times (T_{rb} - T_{rhit}) + 
          R_{rmiss}\times (T_{rb} - T_{rmiss}) + \\
          & R_{w}\times (T_{wb} - T_{wcached})
  \end{split}
\end{equation*}
\vspace{-4mm}

This formula accurately calculates caching profits using real-time 
statistics when caching is enabled, but it malfunctions 
when caching is disabled because $R_{rhit}$ drops to 0.
To overcome this limitation, {\sys} uses the \emph{read ratio} to guide 
cache mode selection, as it is available regardless of whether caching 
is enabled. The caching mode switches when the read ratio surpasses 
or falls below a \emph{threshold} matching the break-even point,
where caching benefits equal its overhead ($P=0$).
This threshold is calculated and recorded in cache headers before
caching is disabled for an object, allowing it to be re-enabled later.

\begin{figure}[t]
  \begin{minipage}{1.\linewidth}
    \centering\includegraphics[scale=.52]{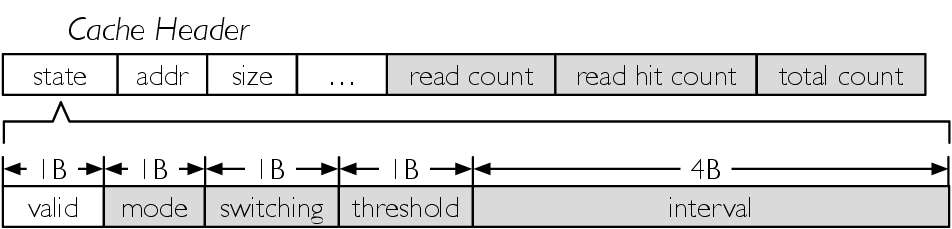}
  \end{minipage} \\[-0pt]
  \begin{minipage}{1.\linewidth}
    \caption{\small{\emph{Additional fields in the cache header
    for fine-grained adaptive caching (marked as grey).
    }}}
    \label{fig:state}
  \end{minipage}  \\[-10pt]
\end{figure}

\stitle{Lightweight statistics.}
To efficiently calculate hit rates and read ratios, {\sys} 
records the counts of reads, read hits, and total operations 
on an object in its cache header, as shown in Fig.~\ref{fig:state}.
These counts are regularly reset to prevent overflow.
For memory consumption concerns, instead of recording operation
latencies of individual objects, {\sys} monitors latencies 
using a thread-local circular buffer for each type of event.
When calculating caching profits, clients average the median latency 
in each thread's buffer.

\subsection{Checking and Switching Cache Modes}\label{sec:select:switch}

{\sys} embeds a \emph{cache mode} in the cache states to determine 
if accesses to the object should go through the cache layer (see 
Fig.~\ref{fig:state}). Cache operations check this mode
after locating the cache header (Lines 1 and 7 in Fig.~\ref{fig:arch}),
and switch the mode when necessary. If the mode is on, caching logic
is executed; otherwise, they fall back to networked accesses.

\stitle{Atomic mode switches.}
Due to the nature of decentralized coherence, each client can
independently switch the cache mode based on local read ratio
statistics. However, clients across different CNs must access 
an object using the same mode to ensure data consistency. 
This requires clients to switch the mode on all CNs atomically. 
An intuitive approach is to acquire a distributed lock before 
switching or checking the cache mode, which is inefficient because 
the mode is checked during every cache operation. 
{\sys} employs an optimistic design that uses locks solely to 
serialize mode switches, allowing mode checks to be lock-free 
via a \emph{switching} flag in cache states (see Fig.~\ref{fig:state}).

\etitle{Mode locks.}
Each object is associated with a CAS-based mode lock
to isolate concurrent switches. 
Each MN stores a fixed number of mode locks. {\sys} hashes
an object's remote address to select its associated mode lock 
on the MN storing its source data.
Hash collisions cause different objects to compete for the
same lock, which restricts the concurrency of mode switches
but does not affect correctness. This is acceptable because
mode switches are infrequent.

\etitle{Switching flags.}
This flag indicates whether the cache mode of an object is being switched. 
Mode checks are paused when the flag is true. Clients set this flag 
to true on each CN before switching modes and reset it afterward,
ensuring that a cache header's mode is consistent across CNs 
when its flag is false.

\stitle{Mode switch interval.}
{\sys} calculates the read ratio threshold and checks whether the
mode should be switched only when the total operation count reaches
a specified \emph{interval} in cache states (see Fig.~\ref{fig:state}).
This design reduces the statistics overhead and limits the
frequency of mode switches. In our implementation, all objects initially 
use 8 operations as the interval to quickly determine the appropriate 
mode. After the first mode switch, the interval is increased to 255 
to enhance the precision of calculated hit rates and read ratios. 

\begin{figure}[t]
  \vspace{1mm}
  \begin{minipage}{1.\linewidth}
    \centering\includegraphics[scale=.44]{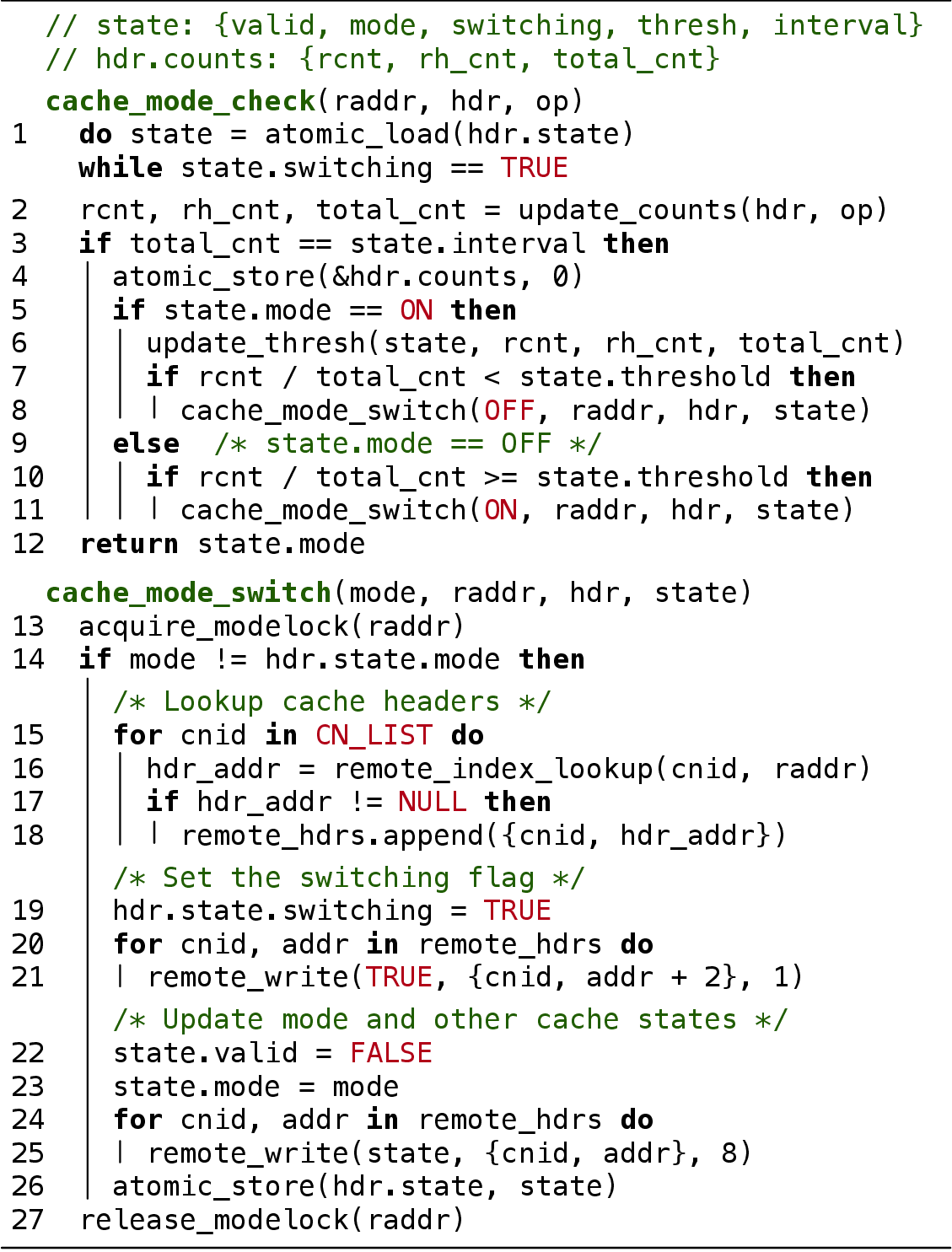}
  \end{minipage} \\[-2pt]
  \begin{minipage}{1.\linewidth}
    \caption{\small{\emph{Pseudo-code of mode check and mode switch.
    }}}
    \label{fig:switch}
  \end{minipage}  \\[-15pt]
\end{figure}

\stitle{Workflow.}
Fig.~\ref{fig:switch} illustrates the workflow of checking
and switching the cache mode, which consists of four steps.

\etitle{Step 1: Check the switching flag.}
The client atomically loads the 8-byte state from the cache header
(Line 1). If the flag is true, indicating an ongoing mode switch,
the client retries loading until the flag becomes false.
This ensures the loaded mode is consistent with the mode on other CNs.

\etitle{Step 2: Update statistics.}
The client updates the count of reads, read hits, and
total operations in the header and acquires their original 
values using atomic fetch-and-add operations (Line 2). 
These operations' addend is determined by the access type 
(\texttt{op}) and the current validity of the header.

\etitle{Step 3: Check switch conditions.}
If the total operation count reaches the mode switch interval (Line 3), 
the client resets the operation counts (Line 4) and compares the read ratio
with the threshold to determine if a mode switch is necessary
(Lines 7 and 10). If caching is enabled, it also updates the threshold
using the caching profit formula (Line 6, see %
\S\ref{sec:select:thresh}).

\etitle{Step 4: Mode switch.}
The client attempts to switch the cache mode if conditions are met.
It acquires the mode lock before starting the switch (Line 13).
After obtaining the lock, the client checks if the mode has already
been switched by a remote client to prevent duplicate switches (Line 14).
If not, it looks up the object's cache headers on all remote CNs 
(Lines 15--18) and sets their switching flags, as well as its own,
to true (Lines 19--21). It then invalidates the cache, updates
the mode, and resets the switching flag for each CN with a remote 
write or atomic store to its cache states (Lines 22--26), 
and releases the lock (Line 27).
The updated threshold and interval are also synchronized to other 
CNs along with other cache states.

\stitle{Default cache modes.}
Mode locks and switching flags ensure mode consistency only
between CNs with the object in their cache index.
For other CNs, when allocating a new cache header for the object, 
the client acquires the mode lock and checks the mode on other CNs
to use as its default mode.
If no CNs have the cache header, the client 
initializes the cache mode as off to avoid caching rarely
accessed objects. The default read ratio threshold for enabling
caching is 75\%.

\section{Implementation}\label{sec:impl}

We implement {\sys} from scratch as a library providing remote 
read and remote write APIs for DM applications, which has roughly 
3,000 lines of C++ code.

\stitle{CN memory usage.}
{\sys} allocates the RDMA memory region for cache data structures
during application startup. It divides the region into fixed-size 
chunks managed in a free list, allocating the minimum number of
chunks necessary to provide sufficient memory for each allocation.
The cache index is initialized with 2 million buckets and corresponding
cache headers, occupying 96\,MB of CN memory, which is negligible 
since CNs have GBs of memory~\cite{fusee,smart}. 

\stitle{Cache eviction.}
When running out of memory during the allocation of cache buffers, 
{\sys} evicts another cached object to reclaim its cache buffer.
While {\sys}'s techniques are orthogonal to cache eviction policies,
it prioritizes objects in the same neighborhood with caching disabled 
as eviction victims. If no such objects are found,
it resorts to adopting conventional policies (e.g., 
least-frequently-used).

\stitle{Dynamic scaling of CNs.}
During invalidations, {\sys} needs to know the IDs of all active 
CNs to collect owners among them. It uses a reliable distributed 
coordinator (e.g., Zookeeper~\cite{zookeeper}) to maintain a list
of CNs running the application, which is cached by each CN.
When the number of CNs changes, the coordinator temporarily
disables caching on all CNs, causing new data accesses to directly 
go through the network, then synchronizes the latest list to them.
If the owner tracking approach changes from broadcast to owner
sets after the change, all cached objects are invalidated and 
all owner sets are cleared to prevent mismatch of existing owners
and owner sets. Caching is re-enabled after this process.

\stitle{Fault Tolerance.}
{\sys} handles node failures by considering cached objects 
and cache metadata on failed nodes as cleared, without recovery.
It detects failures through RDMA operation timeouts. When a timeout
occurs, it requests the coordinator to shut down the destination 
node forcefully, ensuring a consistent view of failures during 
RDMA network partitions. 
Upon detecting a CN failure, the coordinator updates 
the CN list, similar to scaling the number of CNs. 
Upon detecting an MN failure, the coordinator invalidates all
cached objects whose source data was on the failed MN,
as owner sets and mode locks are lost.
Accesses to data on a failed MN returns timeout results for
consistency with network APIs.

\stitle{Batched data accesses.}
Some applications (e.g., transaction engines~\cite{ford,motor}) read or
write several objects in a batch to enhance throughput. {\sys} provides
batched read and write APIs to support these batched data accesses.
In these APIs, network operations for accessing all objects---including 
retrieving data, looking up remote cache indices, and invalidating remote 
caches---are issued in batches.

\stitle{Nested objects.}
{\sys} identifies objects using their remote address, which means
objects must be discrete. Otherwise, writes to one object might 
inadvertently update data belonging to another object without invalidating 
its cached copies.
To support nested objects, {\sys} allows applications to
specify the remote addresses of both the accessed object and its ancestor
object. It looks up the cache index and maintains cache coherence at the
granularity of the ancestor object while retrieving or updating only 
the accessed object. 

\stitle{Atomic operations.}
Atomic operations are unsuitable for decentralized coherence because
the invalidation process is non-atomic. Thus, when the application
executes an atomic operation, {\sys} switches the cache mode of the
accessed object to off and performs a remote atomic operation.

\section{Evaluation}
\label{sec:eval}

\subsection{Experimental Setup}

\nospacestitle{Testbed.}
The experiments were conducted on 10 machines %
connected via a Mellanox 100Gbps switch. Each machine is equipped with 
one 24-core Intel CPU, 128\,GB of RAM, and one ConnectX-4 100Gbps RDMA NIC. 
We use 8 machines as CNs and 1 machine as the MN, following the CN-MN
ratio used in prior work~\cite{smart,ditto,chime}.
An additional machine is reserved as the CN for the centralized manager 
of existing CN-side caches~\cite{polardbsrvless,polardbmp}.
On each CN, 16 cores are dedicated to application clients, with each core 
running a client, and 2\,GB of memory is reserved for caching systems. 
The MN uses 1 core to simulate its wimpy processing power.
Experiments do not involve node failures, which we evaluate separately 
in Appendix B (see our supplementary materials).

\stitle{Applications.}
In addition to evaluating {\sys} with a microbenchmark, we
integrate {\sys} into two real-world DM applications to assess its 
end-to-end improvements. Integration only involves replacing 
the remote access functions with {\sys} APIs, requiring only 
a few dozen LoC.

\etitle{Microbenchmark.}
We built a microbenchmark that performs read and write operations
towards objects stored on MNs, simulating the common data access 
patterns of DM applications. Each object is associated with an
RDMA lock to serialize updates. Read operations retrieve the object
using a remote read and validate its content using versions. 
Write operations acquire the lock, retrieve the object,
write it back using a remote write, and release the lock.
The addresses of all objects are static and known by all clients. 
Each client executes a sequence of operations that is
either generated or specified by 
real-world traces from Twitter~\cite{twittertrace,twittertracegithub}.

\etitle{Database index.}
Sherman~\cite{sherman} is a B$^+$-tree index on DM that serializes tree
modifications with RDMA locks and validates reads with per-node versions.
It represents applications with simple computation but intensive remote data
access. 

\etitle{Transaction engine.}
FORD~\cite{ford} is a DM transaction system that combines two-phase locking and 
optimistic concurrency control to serialize concurrent transactions.
It represents applications with complex computing logic
and data access patterns, such as batched remote operations.

\stitle{Comparing targets.}
We compare {\sys} to not caching data (``{w/o Cache}''), a strategy
used by most DM applications~\cite{sherman,smart,chime,ford,motor,ditto}, and
CN-side cache with a centralized manager (``{CMCache}''), the 
state-of-the-art DM caching approach adopted by 
PolarDB-MP~\cite{polardbmp}. For CMCache, we only present the 
performance of placing the manager on the reserved CN, which consistently 
outperforms placing it on the MN. 
Additionally, we use {\sys} without fine-grained 
adaptive caching (``{DiFache-noAC}'') as a baseline to demonstrate 
the effectiveness of this technique.

\subsection{Data Access Throughput}

\begin{figure}[t]
  \begin{minipage}{1.\linewidth}
    \centering\includegraphics[scale=.42]{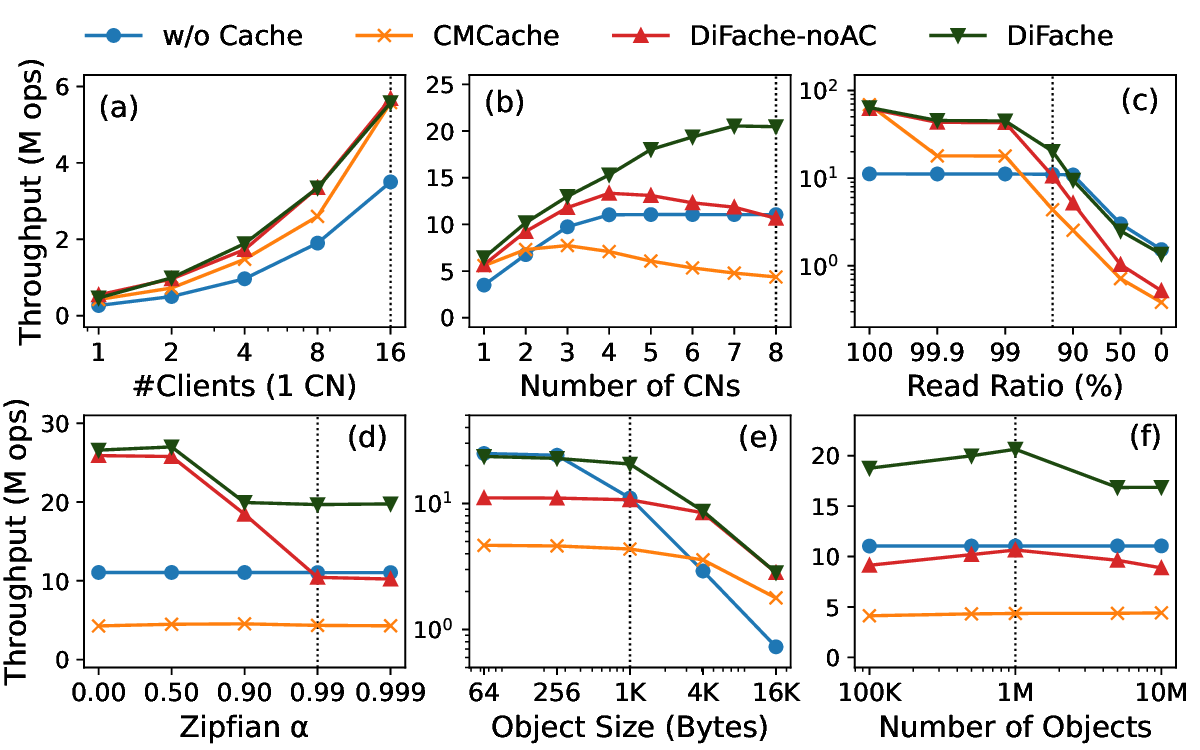}
  \end{minipage} \\[-3pt]
  \begin{minipage}{1.\linewidth}
    \caption{\small{\emph{Throughput of the microbenchmark under
    synthetic workloads with different caching methods and 
    workload parameters. Dotted lines denote the default parameters.
    }}}
    \label{fig:micro}
  \end{minipage}  \\[-5pt]
\end{figure}

\nospacestitle{Synthetic workloads.}
These workloads generate read and write operations with configurable
distributions, where all objects have identical read ratios.
We run the microbenchmark with different workload parameters to
investigate their impact on caching performance.
The default parameters involve 128 clients spread across 8 CNs, 
a 95\% read operation ratio, a Zipfian distribution 
with $\alpha=0.99$, and 1\,KB-sized objects.
These parameters are close to the data access pattern of
DM applications running
read-intensive workloads~\cite{sherman,ford}.
We allocate 1 million objects by default to fix the working set
in the CN-side cache, as cache evictions are not the focus of
{\sys}. Performance results are illustrated in Fig.~\ref{fig:micro}.

\etitle{Number of clients.}
We first increase the number of clients in a single CN from 1 to 
16 (a), and then scale the number of CNs to 8 (b). 
Within 1 CN, all caching methods outperform no caching by up to
1.84\x due to reduced network latency. When scaling to more
CNs, the throughput without caching peaks at 11.05\,M ops
due to MN bandwidth saturation. CMCache reaches peak throughput
at 7.74\,M ops with only 3 CNs, where the centralized manager
becomes fully loaded, leading to throughput drops with more CNs due
to the increased rate of writes and read misses. 
{\sys}-noAC scales linearly with up to 4 CNs via decentralized coherence, 
but beyond that, the growing number of invalidations from
more clients puts significant pressure on the RDMA NIC of each
CN, leading to increased network latencies that degrade throughput.
In contrast, {\sys}'s throughput continues to rise,
outperforming no caching and CMCache by up to 1.86\x and
4.68\x, respectively, as it disables caching for objects that 
generate high CN NIC pressure, mitigating this bottleneck.

\etitle{Read ratio (c).}
All caching methods outperform no caching by up to 6.14\x in
the read-only case because no invalidations are involved.
The throughput of {\sys} and {\sys}-noAC is at least 3.85\x 
higher than no caching with a 99\% or higher read ratio, while 
CMCache is only 1.60\x higher due to extra network
round-trips for requesting the manager during writes and missed
reads. With read ratios of 95\% or less , the throughput of 
{\sys}-noAC and CMCache drops to at least 25.0\% and 33.9\% 
of no caching, respectively, due to higher write latencies.
Oppositely, {\sys} maintains at least the same 
throughput as no caching by
bypassing the cache when accessing write-intensive objects.

\etitle{Skewness (d).}
We vary the Zipfian $\alpha$ to adjust the skewness, with a 
smaller $\alpha$ indicating less skew.
The throughput without caching remains stable despite changes 
in skewness because it does not affect bandwidth consumption.
CMCache also maintains stable throughput as all objects are
cached, resulting in consistent hit rates across different skews.
{\sys}-noAC's throughput drops from 2.33\x 
to 0.93\x of no caching as skewness increases, due to
concentrated invalidations on a few hot objects that
increase CN NIC pressure. In contrast,
{\sys} maintains at least a 1.79\x speedup over no caching
across all skews by selectively disabling caching for 
some objects to mitigate this pressure.

\etitle{Object size (e).}
With object sizes within 256\,B, no caching achieves the
highest throughput since the MN bandwidth is not saturated,
while CMCache and {\sys}-noAC are bottlenecked by serialization
or invalidation costs. 
{\sys} matches the throughput of no caching by disabling
caching for all objects, outperforming {\sys}-noAC by 2.13\x 
and CMCache by 5.07\x. As object size increases, throughput 
without caching drops due to bandwidth saturation,
with {\sys} outperforming it by up to 3.88\x.
The throughput of all caches declines with 4\,KB or larger 
objects because the working set exceeds cache capacity, reducing 
the hit rate.

\etitle{Number of objects (f).}
The throughput of {\sys} and {\sys}-noAC increases slightly
as the number of objects increases to 1\,M because invalidations
are less concentrated and incur smaller CN NIC pressure.
However, it decreases with more objects due to the reduced
hit rates caused by the insufficient cache capacity.
{\sys} outperforms no caching by at least 1.52\x across 
all numbers of objects.

\begin{figure}[t]
  \begin{minipage}{1.\linewidth}
    \centering\includegraphics[scale=.46]{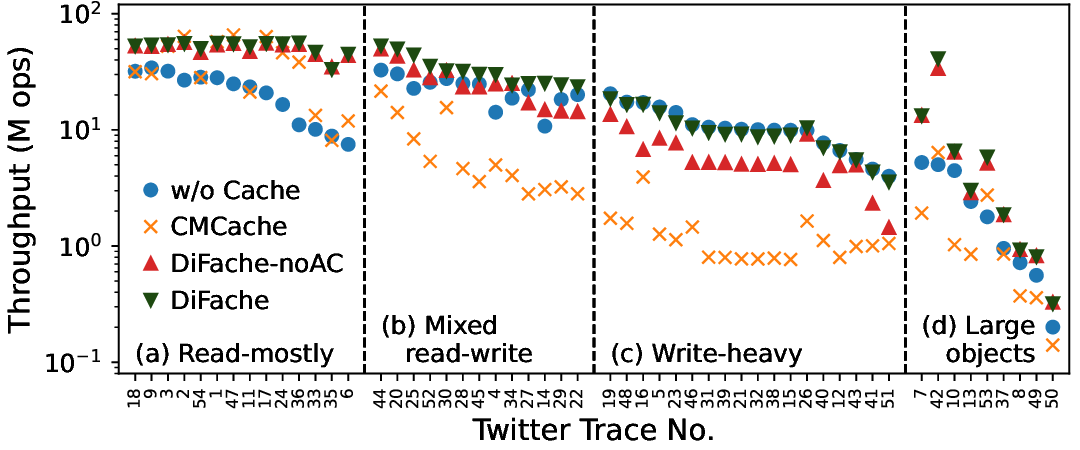}
  \end{minipage} \\[-3pt]
  \begin{minipage}{1.\linewidth}
    \caption{\small{\emph{Throughput of the microbenchmark running
    Twitter traces using different caching methods
    divided into four groups.
    }}}
    \label{fig:twitter}
  \end{minipage}  \\[-10pt]
\end{figure}

\stitle{Real-world workloads.}
We run the microbenchmark with 54 real-world traces from 
Twitter~\cite{twittertrace,twittertracegithub}, where objects
have varying read ratios, sizes, and access frequencies.
All experiments use 128 clients across 8 CNs for maximum load.
When analyzing the results, we group the traces into four
categories with similar performance attributes, as shown in
Fig.~\ref{fig:twitter}.

\etitle{Read-mostly (a).}
When running 14 traces, {\sys} and {\sys}-noAC exhibit 
similar throughput because accesses are mostly reads and 
focus on read-heavy objects.
As the average object size increases, throughput without caching 
drops due to higher MN bandwidth consumption, 
with {\sys} outperforming it by 1.57\x--5.99\x. CMCache performs 
similarly to {\sys} on 5 nearly read-only traces because the manager
is rarely involved, but it is outperformed by 1.20\x--4.08\x on the 
other 9 traces with a lower read ratio.

\etitle{Mixed read-write (b).}
When running 13 traces where objects have varying read ratios, 
CMCache and {\sys}-noAC suffer from high access latencies to 
write-heavy objects, resulting in throughput degradation even 
on traces with a 95\% or higher overall read ratio (No.\,44, 
No.\,20, and No.\,25). 
By adaptively identifying these objects and not caching them, 
{\sys} outperforms CMCache by up to 8.93\x and {\sys}-noAC by
up to 1.70\x. Moreover, it consistently outperforms no caching 
by 1.13\x--2.34\x, even on a trace having only a 65\% overall 
read ratio (No.\,14).

\etitle{Write-heavy (c).}
When running 18 traces, CMCache and {\sys}-noAC experience 
significant throughput degradation by up to 10.83\x and 2.37\x, 
respectively, compared to no caching. This occurs because writes
are frequent or concentrated to a few objects, amplifying
the overhead of caching. In contrast,
{\sys} maintains throughput close to no caching by disabling
caching for almost all objects.

\etitle{Large objects (d).}
When running 9 traces with 2\,KB or larger average object sizes,
the throughput is determined by both object size and read ratio.
The throughput without caching is inversely proportional to the 
object size due to MN bandwidth bottlenecks.
By mitigating this bottleneck, {\sys} and {\sys}-noAC 
consistently outperform no caching by up to 8.16\x, even on
a few traces with only a 50\% read ratio (No.\,10 and No.\,8). 
Conversely, CMCache only outperforms no caching on two traces
with 89\% or higher read ratio (No.\,42 and No.\,53) 
due to its high costs for writes and read misses.

\begin{figure}[t]
  \vspace{2mm}
  \begin{minipage}{1.\linewidth}
    \centering\includegraphics[scale=.42]{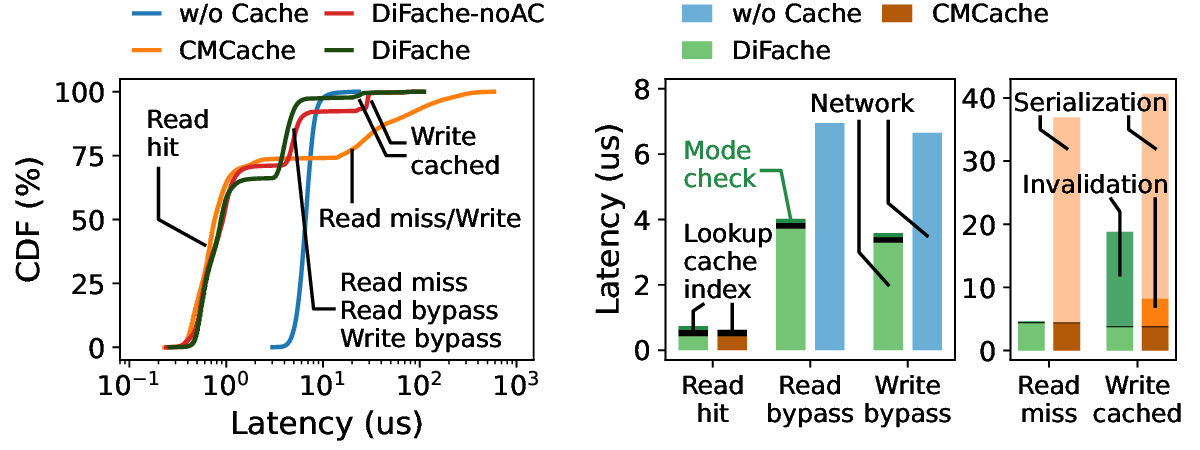}
  \end{minipage} \\[-3pt]
  \begin{minipage}{1.\linewidth}
    \caption{\small{\emph{Latency distribution (left) and median latency breakdown (right) 
	when running the No.\,4 Twitter trace.
    }}}
    \label{fig:brkdown}
  \end{minipage}  \\[-10pt]
\end{figure}

\subsection{Data Access Latency}

We analyze the latency of all five types of data accesses
in {\sys} by comparing their distributions and breakdowns
with other caching methods, as shown in Fig.~\ref{fig:brkdown}.
All systems run a representative Twitter trace (No.\,4) 
using 128 clients across 8 CNs for maximum load.

\stitle{Read hits and read misses to cached objects.}
The median read hit latency of {\sys} and {\sys}-noAC is 0.74\,\us, 
which is 5.7\% higher than CMCache due to cache mode checking. 
However, the read miss latency of CMCache varies from 14.8\,\us to 
585.2\,\us because unhandled cache operations queue at the manager, 
causing significant delays. Oppositely, {\sys} keeps the read
miss latency within 10\,\us (4.6\,\us median) through 
decentralized coherence. 

\stitle{Writes to cached objects.}
Writes in CMCache, like read misses, experience unstable latency due to 
queuing delays. {\sys} addresses this with decentralized invalidation,
reducing median write latency by 53.8\%.
Although {\sys} still requires 14.8\,\us for invalidations due to 
looking up remote cache indexes, it minimizes this overhead by 
decreasing the percentage of operations that trigger invalidations 
(write cached) from 7.7\% to 2.4\% by disabling caching for write-heavy 
objects. This reduces the hit rate by 4.0\% but significantly 
lowers CN NIC pressure and network latency, resulting in a 15.6\% lower 
read miss latency than {\sys}-noAC.

\stitle{Reads and writes bypassing the cache.}
Compared to no caching, {\sys} spends an extra 0.31\,\us on cache index lookup
and mode checking for operations bypassing the cache. However, its latency 
is still 42.2\% lower because it reduces network latency by avoiding MN 
bandwidth saturation.

\begin{figure}[t]
  \vspace{2mm}
  \begin{minipage}{1.\linewidth}
    \centering\includegraphics[scale=.42]{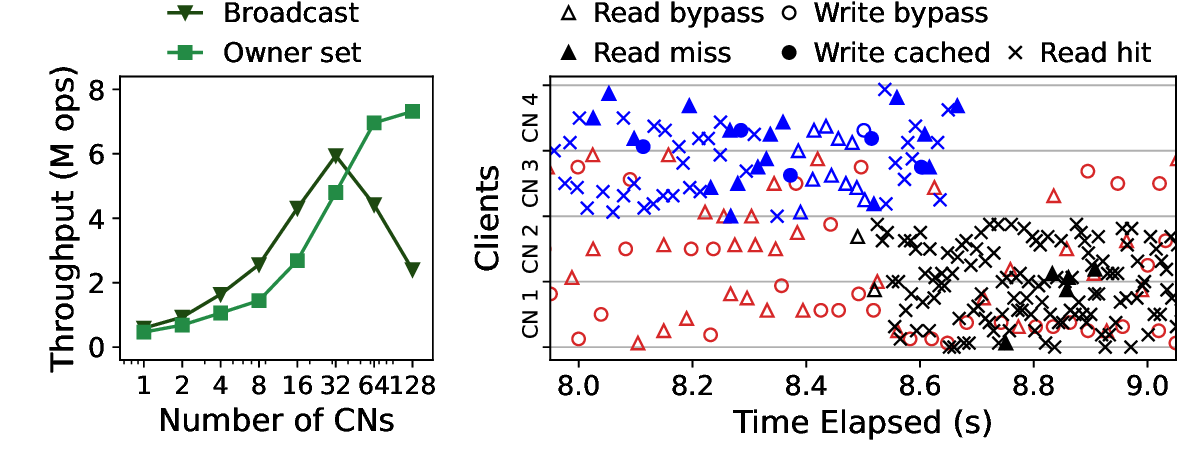}
  \end{minipage} \\[-3pt]
  \begin{minipage}{1.\linewidth}
    \caption{\small{\emph{Throughput scaling of {\sys} using broadcast 
    versus owner sets during invalidations (left), and 
	the read/write accesses for three objects (blue, red, and black) across four CNs
	during one second when running the No.\,22 Twitter trace on 8 CNs (right).
    }}}
    \label{fig:factor}
  \end{minipage}  \\[-10pt]
\end{figure}

\subsection{Owner Tracking Efficiency}

We evaluate the owner tracking efficiency of broadcast versus 
owner sets using the microbenchmark with default parameters.
To simulate more than 8 CNs, each client acts as a virtual 
CN during invalidations. 
As shown in Fig.~\ref{fig:factor} (left), broadcast 
outperforms owner sets by 1.23\x--1.77\x with up to 32 CNs
because it avoids accessing the owner set during read 
misses and writes, resulting in lower cache operation latencies.
However, with over 32 CNs, the throughput of broadcast drops due to 
significant network traffic from invalidation broadcasting, which
degrades remote operation performance.
In contrast, owner sets continue to scale by eliminating 
unnecessary invalidations, outperforming broadcast by 3.05\x at 128 CNs.

\subsection{Cache Mode Switching}

To study the impact of dynamic cache mode switching,
we analyze the cache access patterns in {\sys} when running the
No.\,22 Twitter trace across 8 CNs. 
Fig.~\ref{fig:factor} (right) illustrates the sampled accesses 
(1:200) of three objects on 4 CNs during one second. 
The red object has a stable 50\% read ratio throughout the test, 
so that caching remains disabled for it. 
In contrast, the black object is accessed primarily by reads, which
quickly enables caching and results in a large number of read hits. 
The blue object exhibits dynamic behavior: caching is disabled at 
8.4\,seconds due to frequent writes and read misses %
but is re-enabled 0.1\,seconds later as the read ratio increases again. 
These results demonstrate {\sys}'s ability to adaptively switch
caching modes in response to changing access patterns.

\subsection{Application Performance}

We integrate {\sys} into two real-world DM applications and test
them with representative workloads to demonstrate its end-to-end 
benefits. Both applications use up to 128 clients across 
8 CNs. To estimate CMCache's performance in these applications, we
feed it with the trace of remote data accesses and lock operations 
collected when running the application.

\begin{figure}[t]
  \vspace{1.5mm}
  \begin{minipage}{1.\linewidth}
    \centering\includegraphics[scale=.42]{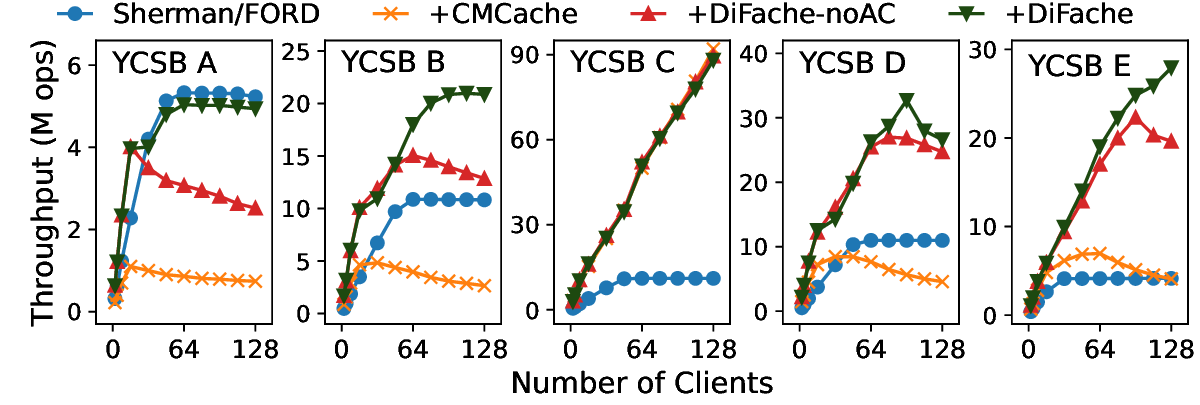}
  \end{minipage} \\[-3.6pt]
  \begin{minipage}{1.\linewidth}
    \centering\includegraphics[scale=.42]{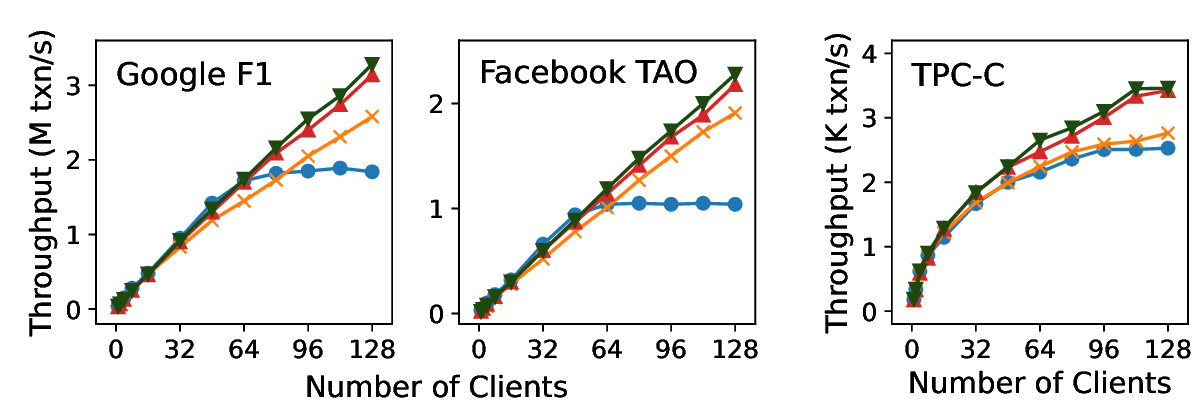}
  \end{minipage} \\[-3pt]
  \begin{minipage}{1.\linewidth}
    \caption{\small{\emph{Throughput scaling of Sherman~\cite{sherman}
    (top) and FORD~\cite{ford} (bottom) under different workloads 
    with different cache schemes.
    }}}
    \label{fig:appperf}
  \end{minipage}  \\[-12pt]
\end{figure}

\stitle{Database index (Sherman~\cite{sherman}).}
We evaluate Sherman using YCSB~\cite{ycsb}, a common benchmark for
evaluating indexes on DM~\cite{smart,chime}.
We use 5 workloads 
with different operation combinations, including 
A (50\% read, 50\% update), B (90\% read, 10\% update), 
C (100\% read), D (95\% read, 5\% insert), and E
(95\% scan, 5\% insert). 

As shown in Fig.~\ref{fig:appperf} (top),
all caching methods scale linearly in the read-only workload (C),
improving Sherman's throughput by up to 7.94\x. However, in other
read-dominated workloads (B, D, and E), CMCache hits a throughput
bottleneck with only 2--4 CNs and degrades
with more CNs due to the centralized manager's saturation.
By adopting decentralized coherence, {\sys}-noAC significantly
improves cache scalability, outperforming Sherman
without caching by 1.38\x (B), 2.46\x (D), and 5.39\x (E) at
peak throughput. However, its throughput decreases after
peaking due to increased invalidation traffic.
{\sys} further enhances scalability by mitigating the impact
of write-heavy objects, improving Sherman's throughput by
1.94\x (B), 2.98\x (D), and 6.73\x (E). In the write-heavy
workload (A), {\sys}-noAC and CMCache experience significant
performance collapse, while {\sys} maintains throughput similar 
to Sherman, as caching is disabled for almost all objects.

\stitle{Transaction engine (FORD~\cite{ford}).}
We evaluate FORD using TPC-C~\cite{tpcc} and two real-world workloads:
F1 from Google~\cite{googlef1} and TAO from Facebook~\cite{facebooktao}, 
aligning with prior work~\cite{ncc}.
For TPC-C, 8 warehouses are allocated to represent high-contention and
compute-intensive workloads. In contrast, F1 and TAO synthesize 
low-contention, data-intensive workloads with 99\% of transactions 
being read-only and having large batch sizes---up to 10 
for F1 and 1,000 for TAO.

As shown in Fig.~\ref{fig:appperf} (bottom), under data-intensive workloads,
FORD reaches peak throughput at 5 CNs due to MN bandwidth saturation. 
In contrast, {\sys} scales linearly to 8 CNs through efficient 
caching, improving FORD's throughput by up to 1.78\x (F1) and 2.19\x 
(TAO). {\sys}-noAC achieves similar throughput due to scarce writes, while
CMCache scales slower due to higher read miss latency, with {\sys} 
outperforming it by 1.27\x (F1) and 1.19\x (TAO). For TPC-C, FORD 
is bottlenecked by lock contention rather than bandwidth,
reducing the impact of caching. Nevertheless, {\sys} still enhances FORD's
peak throughput by 1.37\x due to reduced read latency, whereas CMCache
only improves it by 1.09\x because of inefficient read misses.

\section{Related Work}

\nospacestitle{Disaggregated memory applications.}
Many applications have been adapted to DM architecture
for its resource elasticity, including
key-value stores~\cite{clover,dinomo,fusee,ditto,race}, 
OLTP databases~\cite{dsmdb,polardbmp,ford,rtx,motor},
and tree-based indexes~\cite{sherman,smart,rolex,marlin,chime}.
Most of them only cache small metadata, %
leaving data objects uncached to avoid expensive coherence. 
This often leads to high access latency and MN bandwidth
saturation, which {\sys} addresses through coherent caching.
DINOMO~\cite{dinomo} caches data in addition to metadata but 
prohibits different CNs from updating the same data, 
limiting computing resource elasticity.
PolarDB-MP~\cite{polardbmp} maintains cache coherence using 
centralized managers on MNs, which become bottlenecks 
as the client number increases. {\sys} enhances
scalability using a decentralized coherence design.

\stitle{CN-side caching approaches.}
Several approaches have been proposed to accelerate remote data 
access by caching data in CN memory, 
either at the page granularity~\cite{infiniswap,fastswap,canvas} 
or object granularity~\cite{aifm,kona,atlas}. 
Unlike {\sys}, these approaches lack cross-CN cache coherence, 
which prevents DM applications from scaling correctly across multiple CNs. 
MIND~\cite{mind} implements coherence using distributed shared memory~\cite{ivy,gam}, 
and uses a programmable switch as a centralized manager. %
Even so, it has limited scalability since it treats all local memory accesses 
as cache accesses, ignoring DM application semantics.
In contrast, {\sys} only caches explicit remote accesses,
significantly reducing invalidations.
SELCC~\cite{selcc} ties coherence maintenance to lock operations, 
requiring all remote accesses to use reader-writer locks, 
which excludes DM applications with lock-free reads.
{\sys} supports both lock-based and lock-free DM applications.

\section{Conclusion}

This paper presents {\sys}, an efficient and scalable CN-side cache
for DM applications. Evaluation using real-world traces and
applications confirms the benefits of {\sys}.

\nocite{cxlamd,pasha}
% The 'abbrvnat' bibliography style is recommended.

\balance

\small{
  \bibliographystyle{plain}
  \bibliography{bib/dblp,bib/misc}
}

\clearpage

\begin{appendix}

\appendix
\renewcommand{\thesection}{Appendix \Alph{section}}

\section{Discussion of CXL-based DM}

While this paper focuses on applications and distributed cache designs 
in RDMA-based DM environments, efficient software CN-side caching is also 
crucial for building efficient and scalable applications 
in CXL-based DM environments.
Although upcoming CXL 3.0 products support hardware cache coherence, 
this coherence is typically limited to a small memory region---at most 
a few hundred MBs---because supporting cache coherence for the entire
memory space is too costly~\cite{cxlamd,pasha}. Accesses to remote memory outside
the hardware-coherent region still rely on remote memory access (RMA) 
APIs~\cite{cxlamd}. 
By replacing these remote accesses with cache accesses, {\sys} can 
accelerate applications on CXL-based DM as well as those designed 
for RDMA-based DM.

The hardware-coherent region in CXL-based DM environments offers
opportunities for {\sys} to realize more efficient decentralized invalidation.
Specifically, the cache index and cache headers can be placed in the 
hardware-coherent region and shared among all CNs, with each object's owner 
set embedded in its cache header. This approach eliminates the need to look 
up remote cache header addresses on each owner and saves the extra lookup
for locating the owner set, enabling a more scalable invalidation design.
Cache mode switches also become less expensive, as they only require
updating the mode in the shared cache header instead of broadcasting
mode updates to all CNs. We leave these designs on CXL-based platforms
as future work.

\begin{figure}[t]
  \vspace{2mm}
  \begin{minipage}{1.\linewidth}
    \centering\includegraphics[scale=.42]{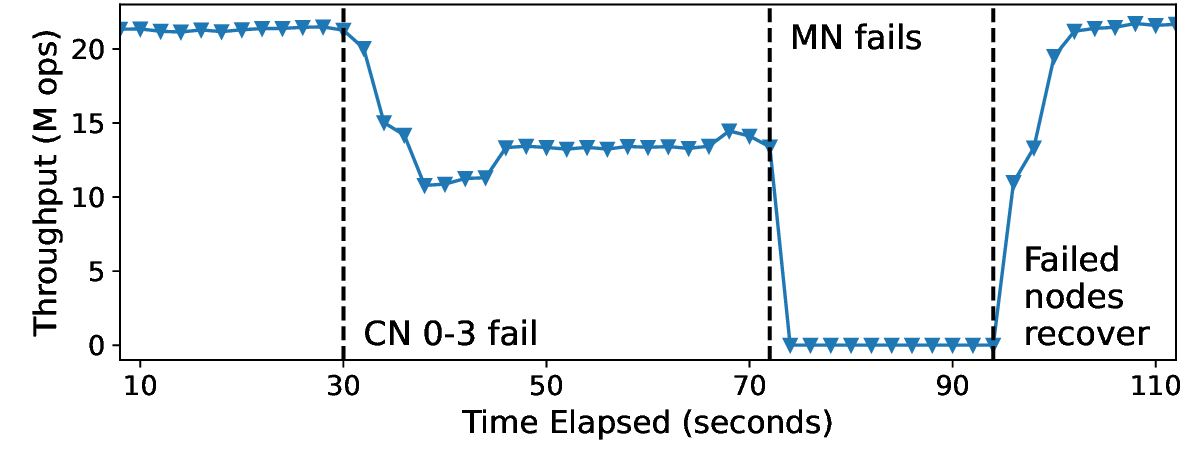}
  \end{minipage} \\[2pt]
  \begin{minipage}{1.\linewidth}
    \caption{\small{\emph{The timeline of {\sys}'s throughput
    with the presence of CN and MN failures. 
    }}}
    \label{fig:failure}
  \end{minipage}  \\[-5pt]
\end{figure}

\section{Fault Tolerance Experiment}

To investigate {\sys}'s behavior during node failures, we collect a timeline of
its throughput as node failures and recoveries occur while running the 
microbenchmark with default parameters, as shown in 
Fig.~\ref{fig:failure}. Node failures are simulated by killing the {\sys} 
process on the affected nodes. We first kill four CNs (0--3) in sequence, 
then kill the MN, and finally recover all of the previously killed nodes.

After detecting CN failures through timed-out invalidations, {\sys} 
temporarily disables caching and synchronizes the updated CN list to the
surviving CNs (see details in \S\ref{sec:impl}). 
This causes throughput to drop to a level similar to that 
without caching, between 38 and 44\,seconds. 
Throughput rises again once synchronization is complete and caching
is re-enabled. When the MN fails, system throughput drops to zero 
because all remote data accesses time out. While waiting for the MN
to recover, {\sys} invalidates caches on all CNs.
Throughput returns to peak levels within 8\,seconds after the failed nodes 
recover, as the cache is refilled and cache modes are adapted
to the workload pattern.

\end{appendix}

\end{document}